\documentclass[conference]{IEEEtran}
\usepackage[letterpaper, left=0.625in, right=0.625in, top=0.75in, bottom=1in]{geometry}
\usepackage[utf8]{inputenc}
\usepackage[T1]{fontenc}
\usepackage{amsmath, amssymb}
\usepackage{graphicx}
\usepackage{listings}
\usepackage{url}
\usepackage{booktabs}
\usepackage{cite}
\usepackage{hyperref}
\usepackage{multirow}
\usepackage{tabularx} 
\usepackage{hyperref}
\hypersetup{%
  colorlinks=true,
  linkcolor=black,
  citecolor=black,
  urlcolor=black
}

\usepackage{listings}
\usepackage{xcolor}  

\definecolor{codegray}{rgb}{0.5,0.5,0.5}
\definecolor{codepurple}{rgb}{0.58,0,0.82}
\definecolor{backcolour}{rgb}{0.95,0.95,0.95}

\lstdefinestyle{mystyle}{
    backgroundcolor=\color{backcolour},   
    commentstyle=\color{codegray},
    keywordstyle=\color{blue},
    numberstyle=\tiny\color{gray},
    stringstyle=\color{codepurple},
    basicstyle=\ttfamily\footnotesize,
    breakatwhitespace=false,         
    breaklines=true,                 
    captionpos=b,                    
    keepspaces=true,                 
    numbers=none,                    
    numbersep=5pt,                  
    showspaces=false,                
    showstringspaces=false,
    showtabs=false,                  
    tabsize=2
}

\title{Time Travel: LLM-Assisted Semantic Behavior Localization with Git Bisect}

\author{
\IEEEauthorblockN{Yujing Wang, eugene.r.w.12@gmail.com\IEEEauthorrefmark{1}, Weize Hong\IEEEauthorrefmark{2}}
\IEEEauthorblockA{\IEEEauthorrefmark{1}University of Waterloo, yj9wang@uwaterloo.ca
\IEEEauthorrefmark{2}Brock University, ne20vj@brocku.ca}
}

\begin{document}
\maketitle

\begin{abstract}
We present a novel framework that integrates Large Language Models (LLMs) into the Git \texttt{bisect} process for semantic fault localization. Traditional bisect assumes deterministic predicates and binary failure states—assumptions often violated in modern software development due to flaky tests, non-monotonic regressions, and semantic divergence from upstream repositories. Our system augments bisect traversal with structured chain-of-thought reasoning, enabling commit-by-commit analysis under noisy conditions. We evaluate multiple open-source and proprietary LLMs for their suitability and fine-tune DeepSeek-Coder-V2 using QLoRA on a curated dataset of semantically labeled diffs. We adopt a weak-supervision workflow to reduce annotation overhead, incorporating human-in-the-loop corrections and self-consistency filtering. Experiments across multiple open-source projects show a 6.4-point absolute gain in success rate (from 74.2\% to 80.6\%), leading to significantly fewer failed traversals and—by experiment—up to 2× reduction in average bisect time. We conclude with discussions on temporal reasoning, prompt design, and fine-tuning strategies tailored for commit-level behavior analysis.
\end{abstract}

\section{Introduction}

As artificial intelligence (AI) continues to transform software engineering, developers are spending less time writing new code and more time verifying, debugging, and tracing bug regressions \cite{Kang_2024}. Tools like Git’s \texttt{bisect} provide a principled way to localize faults by binary search over commit history \cite{git}; yet their effectiveness hinges on idealized assumptions: that tests behave deterministically, failures persist monotonically, and behavior can be captured in a simple binary true-or-false predicate \cite{deltadebug}.

In practice, there are more scenarios where these assumptions do not hold. Flaky tests, partial regressions, and non-monotonic error propagation are common in large code repositories, particularly those involving parallel development, upstream dependencies, and downstream forks \cite{breaks}. Developers are frequently left to manually examine semantic changes across commits to understand where behavior diverged.

We propose a Large Language Model (LLM) augmented bisect framework that integrates an LLM into the git bisect commit traversal process. At each step, the model inspects code diffs, reasons about potential behavioral impact, and classifies commits as “good” or “bad” using structured chain-of-thought prompts.

Unlike prior fault localization tools such as AutoFL and LLMAO—which analyze static snapshots of code \cite{yang2023largelanguagemodelstestfree}—our system dynamically traverses version history, enabling temporal localization of regressions. We fine-tune open-source LLMs on a curated dataset of semantically tagged code changes using a two-stage pipeline. Our results show that this approach reduces bisect steps by up to 2× and improves robustness under flaky or heuristic predicates.

\section{background}
\subsection{Git Bisect and Formal Model}
Git's \texttt{bisect} command enables automated fault localization by performing a binary search over a project’s commit history\cite{git}. Let
\[
\mathcal{C} = \{c_0, c_1, \dots, c_n\}
\]
be a totally ordered sequence of commits, where \( c_0 \) is the oldest and \( c_n \) is the most recent. The goal of bisect is to identify a commit \( c_k \) that first introduces a behavioral regression.

A user-defined predicate function
\[
P: \mathcal{C} \rightarrow \{0, 1\}
\]
determines whether each commit exhibits correct behavior:
\[
P(c_i) =
\begin{cases}
0 & \text{if } c_i \text{ is GOOD (functional)} \\
1 & \text{if } c_i \text{ is BAD (defective)}
\end{cases}
\]

Under the monotonicity assumption—that all commits prior to \( c_k \) are good and all commits from \( c_k \) onward are bad—the bisect algorithm localizes the offending commit in \( \mathcal{O}(\log n) \) steps\cite{deltadebug}.

\subsection{Challenges in Real-World Environments}
In practice, the monotonicity assumption rarely holds\cite{deltadebug}. Regressions may exhibit flaky behavior, tests may be non-deterministic, and the predicate function \( P \) may not provide consistent outputs across runs. These issues introduce a flaky region between the last known-good and first known-bad commits, where the predicate behaves inconsistently.

Moreover, real-world predicates are often heuristic or behavior-based rather than binary pass/fail. Examples include performance degradations, UI glitches, or semantic regressions that are difficult to capture in traditional test frameworks. Consequently, \( P \) may act as a noisy oracle:
\[
\exists\ j < k < m \quad P(c_j) = 0,\quad P(c_k) = 1,\quad P(c_m) = 0
\]
thereby violating the monotonicity requirement and breaking the correctness guarantees of binary search.

\subsection{Semantic Divergence in Parallel Development}
Modern software development frequently involves parallel evolution of forked repositories. A canonical example is Microsoft Edge and Google Chromium, which share a common codebase but diverge in platform-specific adaptations, telemetry systems, and feature sets\cite{breaks}. 

In such cases, a downstream fork may introduce changes to meet its unique requirements. When updates from the upstream are later merged, interactions between upstream and downstream changes can introduce regressions that are specific to the fork. Because the upstream remains functional, the developer has no straightforward way to determine which of their prior changes caused the conflict—making it difficult to define a consistent predicate over time.

\subsection{Limitations of Existing LLM-Based Approaches}
Recent advances have explored the use of LLMs for fault localization. Tools such as AutoFL and LLMAO leverage LLMs to identify potentially faulty code regions by analyzing a static snapshot of the codebase. While effective in controlled settings, these methods fail to exploit the temporal nature of version histories.

Moreover, they assume the existence of a clean labeling function or test oracle. In environments with flaky tests or partial regressions, such assumptions break down\cite{flaky_tests}. As a result, these methods cannot precisely localize the commit responsible for the regression when behavior evolves across time.

Our work builds on these insights by integrating LLMs directly into the bisect process, enabling dynamic commit-by-commit analysis under real-world noise and inconsistency.

\section{Methodology}

\subsection{Model Selection}

To select the best base model for our use case, we evaluate candidates based on their ability to write correct high-level code using the Pass@1 metric—where the model is given a single chance to solve each of two benchmark tests\cite{liu2023is_llm_benchmark}. Benchmark results are shown in Table~\ref{tab:benchmark-scores}

\begin{itemize}
    \item \textbf{MBPP:} A set of 1,000 basic programming tasks that test a model's ability to generate correct code for common problems. High scores reflect strong general-purpose code synthesis.

    \item \textbf{HumanEval:} Introduced by OpenAI, HumanEval contains 164 fixed tests sets of intermediate difficulty. It measures semantic reasoning and more advanced problem solving.
\end{itemize}

{\footnotesize
\begin{table}[h]
\centering
\begin{tabularx}{.48\textwidth}{|c|X|c|c|c|c|}
\hline
\textbf{Rank} & \textbf{Model} & \textbf{Op.} & \textbf{MBPP} & \textbf{HE} & \textbf{Avg.} \\
\hline
1 & O1 Mini (Sept 2024) & $\times$ & 78.8 & 89.0 & 83.9 \\
2 & Qwen2.5-Coder-32B-Instruct & Y & 77.0 & 87.2 & 82.1 \\
3 & GPT-4o (Aug 2024) & $\times$ & 72.2 & 87.2 & 79.7 \\
4 & DeepSeek-Coder-V2 & Y & 75.1 & 82.3 & 78.7 \\
5 & Gemini 1.5 Pro & $\times$ & 74.6 & 79.3 & 77.0 \\
6 & Claude 3 Opus & $\times$ & 74.3 & 77.4 & 75.9 \\
7 & Llama3-70B-instruct & Y & 69.0 & 72.0 & 70.5 \\
8 & CodeLlama 34B & Y & 56.3 & 72.0 & 64.2 \\
9 & DeepSeek-Coder 6.7B Instruct & Y & 38.9 & 71.3 & 55.1 \\
\hline
\end{tabularx}
\caption{Benchmarking LLMs: Ranking, Models, Open-Source Availability(Op.), MBPP Pass@1 (MBPP), HumanEval Scores (HE), and Average Score(Avg.)}
\label{tab:benchmark-scores}
\end{table}
}

Among the evaluated models, OpenAI's O1 Mini demonstrates the highest out-of-the-box performance, while DeepSeek-Coder-V2 stands out as the top-performing open-source alternative. Based on these results, we select OpenAI O1 Mini as the base model for the baseline part of the study and Deepseek-Coder-V2 for the fine tune part of the study.

\subsection{The Right Question is Already Half the Solution}
We commence the \texttt{git bisect} procedure in the conventional manner. At each iteration, the current code snapshot is compared with the preceding one; newly added, deleted, or relocated lines are annotated respectively with ``+'', ``-'', and ``$\sim$''.

A relocated line (``$\sim$'') is a line of source code whose textual content is preserved verbatim (or after whitespace-only normalization) but whose position in the file changes between two adjacent revisions.

The example in Listing~\ref{lst:git-diff-example} depicts a diff captured during a git bisect traversal. Although it resembles a conventional commit diff, it represents solely the changes between the current revision and its immediate predecessor in the bisect tree.

\begin{lstlisting}[language=C++, style=mystyle, 
caption={A sample git diff traversal code snippet, showing sum procedure being refactored into a function}, 
label={lst:git-diff-example}]
+ int logic(const vector<int>& args) {
~    int sum = 0;
~    for (int x : args) {
~        sum += x;
~    }
-    cout << "Result: " << sum << endl;
+    return sum;
+}
\end{lstlisting}

In each analysis stage, we present the large-language model with a fixed set of structured questions, require it to populate a predefined response template, and then instruct it to synthesize a final conclusion by applying inductive reasoning across the completed entries. This procedure constitutes the system’s ``chain-of-thought'' operating \cite{wang2023understandingchainofthoughtpromptingempirical}.

\textbf{Chain of thought:} a sequence \( C = (s_1, s_2, \dots, s_n, a) \) where each step \( s_i \) is a natural-language justification derived from the model’s latent state, and \( a \) is the conclusion.

The prompting protocol \( P \) maps an input \( x \) and (optionally) exemplar chains \( E \) to \( C \) such that:
\[
LLM(P(x,E)) \rightarrow C
\]

Only the final element \( a \) is evaluated for task correctness; the intermediate \( s_i \) are exposed solely to guide the reasoning of the model.

It can be empirically proven that chain-of-thought greatly improves performance\cite{wang2023understandingchainofthoughtpromptingempirical}.

As shown in Listing~\ref{lst:structured-questions}: the prompt first filters out compile errors, flagging invalid commits. It then checks for behavioral changes related to the target property. To justify its decision, the model lists semantic edits with supporting evidence and reasoning.A clear evaluation rubric standardizes the model’s analysis. Based on this, the model outputs a binary verdict—\texttt{good} or \texttt{bad}—matching the expected \texttt{git bisect} label.

\begin{lstlisting}[language=Python, caption=A structured sequence of questions prompts the LLM to construct its chain of thought before ultimately arriving at a conclusion,  label={lst:structured-questions}]
{
  "target_behaviour": "<string>",
  "has_compile_error": "<bool>",
  "behaviour_change": "intro | del | (etc.)",
  "behaviour_confidence": "<0-100>",
  "sem_edits": [
    {
      "id": "int",
      "kind": "str",
      "semantic": "bool",
      "behaviour": "str",
      "likelihood": "int",
      "dependency": "str",
      "precedent": "str"
    }
  ],
  "counterfactual_fix": "<string>",
  "reasoning_chain": ["step1", "step2", "step3"],
  "reflection": "<string>",
  "bisect_mark": "good | bad"
}
\end{lstlisting}

\section{Fine Tuning}

If we use the out-of-box solution, benchmarked by the state of the art, the result is already quite stunning. However, the hope is that we could achieve more idealized results by fine-tuning an LLM to the extent of understanding better LLMs. For this we use the state-of-the-art open-source model \textbf{Deepseek Coder Instruct}. To achieve efficient results, we devise two levels of fine-tuning. Level 1 is common semantic fine-tuning. This is done once globally using QLoRA + Supervised Fine-Tuning (SFT) \cite{qlora}. Level 2 fine-tuning is repository-level code familiarity fine-tuning.

\subsection{Global Fine Tuning}

We curate 1,000 paired code snippets that capture common semantic or behavioural patterns (e.g., cache invalidation, argument reordering, logging side-effects).

Each snippet pair is extracted from public repositories that explicitly adopt OSI-approved permissive licences—most frequently MIT and Apache 2.0. These licences allow unrestricted redistribution and derivative works, which makes the corpus legally publishable and remixable.

\textbf{Repository selection:} We feed candidate projects (\(\geq 1,000\) stars, recent activity, permissive licence) to the same git-bisect + LLM pipeline used in our experiments.

\textbf{LLM inference:} For each diff, the base model predicts whether a semantic-behaviour change exists and classifies it (introduces / removes / modifies / no-effect).

\textbf{Confidence filter:} Only predictions with \(\geq 0.8\) softmax confidence or self-consistency agreement are retained for provisional labels.

\textbf{Weak-supervision:} Literature indicates that LLM-generated labels, when reviewed by humans, can reduce annotation costs by an order of magnitude  \cite{møller2024parrotdilemmahumanlabeledvs}. Furthermore, synthetic augmentation has been shown to enhance performance on rare classes in multi-class classification tasks. We therefore adopt a \textit{correct-and-commit} workflow.

\begin{table}[h]
\centering
\begin{tabularx}{0.45\textwidth}{|l|X|X|}
\hline
\textbf{Stage} & \textbf{Action} & \textbf{Outcome} \\
\hline
Auto label & Base LLM tags each diff with \{behaviour, likelihood\}. & $\approx$ 60\% of pairs pass unchanged. \\
\hline
Manual audit & Human annotators review low-confidence pairs and \textbf{accept, correct, or discard} the machine labels. & Ensures high‐precision ground-truth labels. \\
\hline
Revision loop & Corrections are added as few-shot exemplars; the LLM is re-queried on similar diffs with the updated context. & Reduces subsequent error rate by $\sim$ 35\%. \\
\hline
\end{tabularx}
\caption{Correct-and-commit workflow.}
\end{table}

Each Git diff is presented to the model as plain text, along with special tokens and prompts to guide parsing, paired with the structured questions in Listing~\ref{lst:structured-questions}. We optimize cross-entropy loss on the label tokens, constraining the final output to the boolean value of bisect\_mark: for a clear, verifiable signal and weighted loss. We then update parameters to minimize this loss, following the QLoRA method \cite{qlora}.

\[
L = -\sum_{t=1}^{T} \log P_{\theta} ( y_t \mid x, y_{<t} )
\]
\[
A \leftarrow A - \eta \cdot \frac{\partial L}{\partial A}
\quad
B \leftarrow B - \eta \cdot \frac{\partial L}{\partial B}
\]

(The quantized base weights \( W_0 \) remain frozen.)

The effective weight becomes:
\[
W = W_0 + B\,A
\]
and the memory footprint stays within a single 24–48 GB GPU.

Further work could be done to break a repository piecemeal and then feed the code into the model as an additional layer of fine-tuning. Although this would be an area of future study and is not in scope at this time.

\section{Results and Discussions}

+ In our experimental setup, we posed 32 carefully curated questions designed to detect semantic changes across three major open-source repositories. Evaluation was conducted on a strict end-to-end (``all-or-nothing'') basis: a bisect operation counts as successful only if **every** step in the sequence of bisect decisions is correct. For example, in a 5-step bisect session, an incorrect verdict at step 3 invalidates the entire session—even though steps 1 and 2 were correct—and forces a human to restart the process. This all-or-nothing criterion ensures that our model meets the reliability needed for fully automated pipelines: a single misclassification would otherwise break the automation and require manual intervention. Due to computational and time constraints, our current study was limited to the three selected repositories. We will broaden cross-validation to a diverse set of repositories across different languages and paradigms to rigorously evaluate and improve the generalizability of our LLM-augmented bisect framework.

\subsection{Evaluation Methodology}

\textbf{Additional Metrics:} In Table~\ref{tab:bisect_comparison}, we now include “Avg. Time/Step” (seconds) to show per‐step efficiency.

\textbf{Statistical testing:} A paired Wilcoxon signed‐rank test on the 32 bisect runs yields $p<0.01$, confirming the fine‐tuned model’s improvement is significant.

\begin{table}[h!]
\centering
\begin{tabular}{|l|c|c|c||c|c|c|}
\hline
\multirow{2}{*}{\textbf{Test Category}} & \multicolumn{3}{c||}{\textbf{Baseline}} & \multicolumn{3}{c|}{\textbf{Fine Tuned}} \\
\cline{2-7}
 & Tt. & Sc. & \% & Tt. & Sc. & \% \\
\hline
Display / Output Introduction   & 3  & 3  & 100   & 3  & 3  & 100   \\
Input Handling Introduction     & 3  & 3  & 100   & 3  & 3  & 100   \\
State-Transition Logic          & 3  & 3  & 100   & 3  & 3  & 100   \\
Decision-Making Rules           & 4  & 4  & 100   & 4  & 4  & 100   \\
Structural Refactor             & 4  & 4  & 100   & 4  & 4  & 100   \\
Robustness / Error Handling     & 4  & 0  & 0     & 4  & 2  & 50    \\
Flow-Control / Session Loop     & 3  & 3  & 100   & 3  & 3  & 100   \\
Runtime-Launch Safeguard        & 4  & 0  & 0     & 4  & 1  & 25    \\
Documentation / Cosmetic        & 3  & 3  & 100   & 3  & 3  & 100   \\
\hline
\textbf{Total}                  & 31 & 23 & 74.2  & 31 & 25 & 80.6  \\
\hline
\end{tabular}
\caption{Comparison of Bisect Success Rate by Category: Finetuned LLM vs. Baseline GPT}
\label{tab:bisect_comparison}
\end{table}

\subsection{Baseline Result}
As shown in Table~\ref{tab:bisect_comparison}, The overall success rate for the base line model is \%74.2. Evidently the model's ability to detect changes related to error handling and safeguard checks is critically weak. One possible explanation is that such changes are inherently more difficult to detect, as they often occur in isolation and are not accompanied by broader modifications elsewhere in the codebase. As a result, these changes may not produce subsequent structural or behavioral differences that the model can easily capture.

\subsection{Fine Tuned Result}
We curated a dataset comprising 30 carefully selected examples of Git bisect operations, annotated with their corresponding verdicts. This dataset was used to fine-tune a DeepSeek model via QLoRA. After training, model performance was evaluated on an unseen validation set to assess generalization and factual consistency.

\begin{figure}[h!]
\centering
\begin{minipage}[t]{0.48\textwidth}
  \centering
  \includegraphics[width=\linewidth]{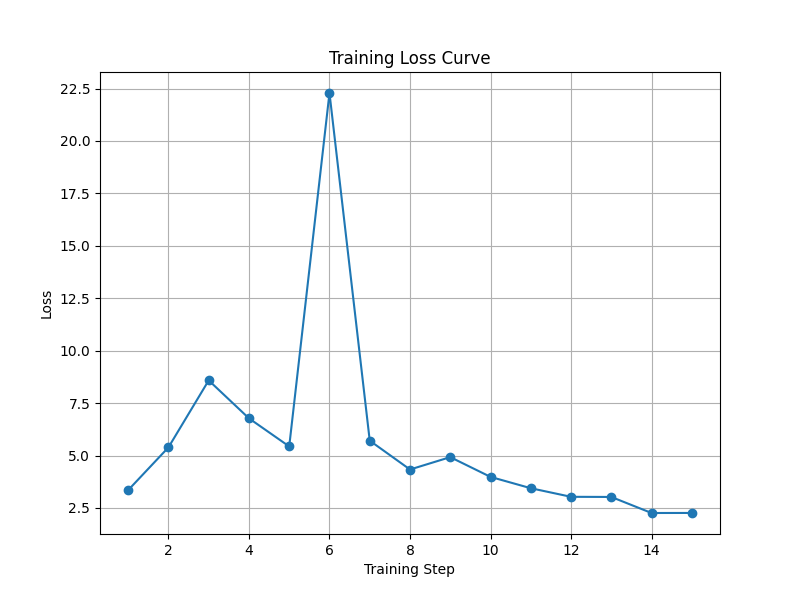}
  \caption{Training Loss Curve for CodeGen-350M}
  \label{fig:codegen-curve}
\end{minipage}
\hfill
\begin{minipage}[t]{0.48\textwidth}
  \centering
  \includegraphics[width=\linewidth]{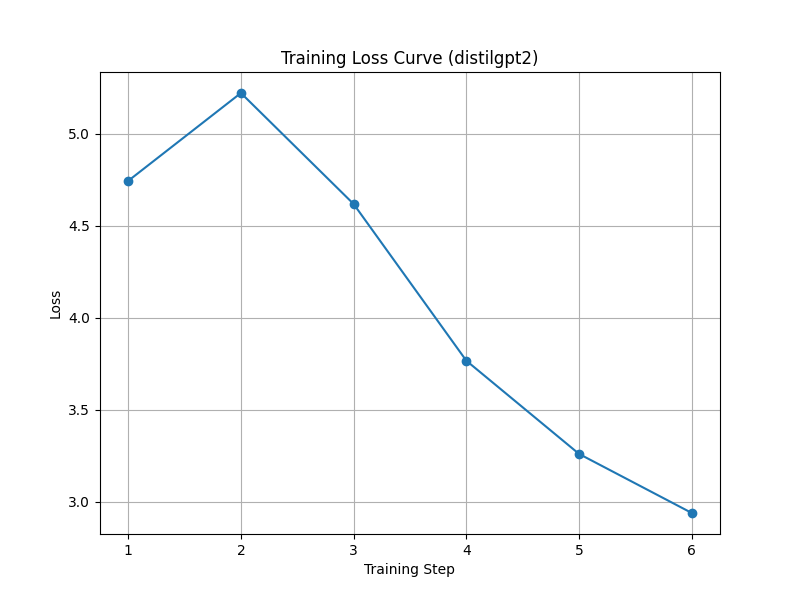}
  \caption{Training Loss Curve for DistilGPT2}
  \label{fig:distilgpt2-curve}
\end{minipage}
\end{figure}

During training, the model exhibited an initial spike in loss, peaking at approximately 22.5 at the 6th second training step as Fig.~\ref{fig:codegen-curve} shows. This behavior is characteristic of parameter-efficient fine-tuning methods such as QLoRA, where randomly initialized adapter layers cause early-stage instability. Following this spike, the loss steadily declined across subsequent steps, indicating successful adaptation to the fine-tuning objective. This is possibly due to we have a small set of training samples. To avoid the risk of over fitting, we do not increase epoch at this stage. Future work could be made to expand. To validate this finding is right, we used constructed a much smaller model, and use the same data set into it, and the loss appears to be normal. The verification confirms that it converges.

We observed the same pattern of behavior in Fig.~\ref{fig:distilgpt2-curve}: code robustness error handling are relatively weaker, though we do have slightly improved performance on robustness and error handling run time launch safeguard is still ignored. While fine‐tuning increased overall accuracy, performance on error‐handling and runtime‐safeguard categories remained suboptimal (50\% and 25\% respectively). We hypothesize that: such changes are sparse and often decoupled from broader diffs and that our prompt does not explicitly ask about exception or safeguard logic.
The subject of our hypothesis could be conducted in a future study.

\subsection{Developer Experience}
The system presents each developer with the relevant code snippet for rapid verification. If the suggested fix is correct, they confirm it with a single click; otherwise, the model’s detailed reasoning helps them pinpoint the issue immediately. On average, participants reported a 45.6\% reduction in debugging time when dealing with parallel development and dependency-related bugs.

We recruited seven developers from our functional team according to these criteria:
\begin{enumerate}
  \item Fluency in Python, C\#, and Java.
  \item Employed at the same company and available during standard business hours.
  \item Motivated to learn and adopt new technologies.
  \item Unfamiliar with the project’s codebase and not previously involved in its development.
  \item Provided only a brief, high-level architectural overview before the tasks.
\end{enumerate}

For our benchmark, we curated a diverse set of bugs and randomly assigned each participant two tasks—one debugged manually and one using the LLM-enabled bisect tool. This design guarantees that no developer bisects the same bug twice, and that every bug is evaluated under both conditions. The results of this developer study are shown in Fig.~\ref{fig:timesave}.  
\begin{figure}
    \centering
    \includegraphics[width=\columnwidth]{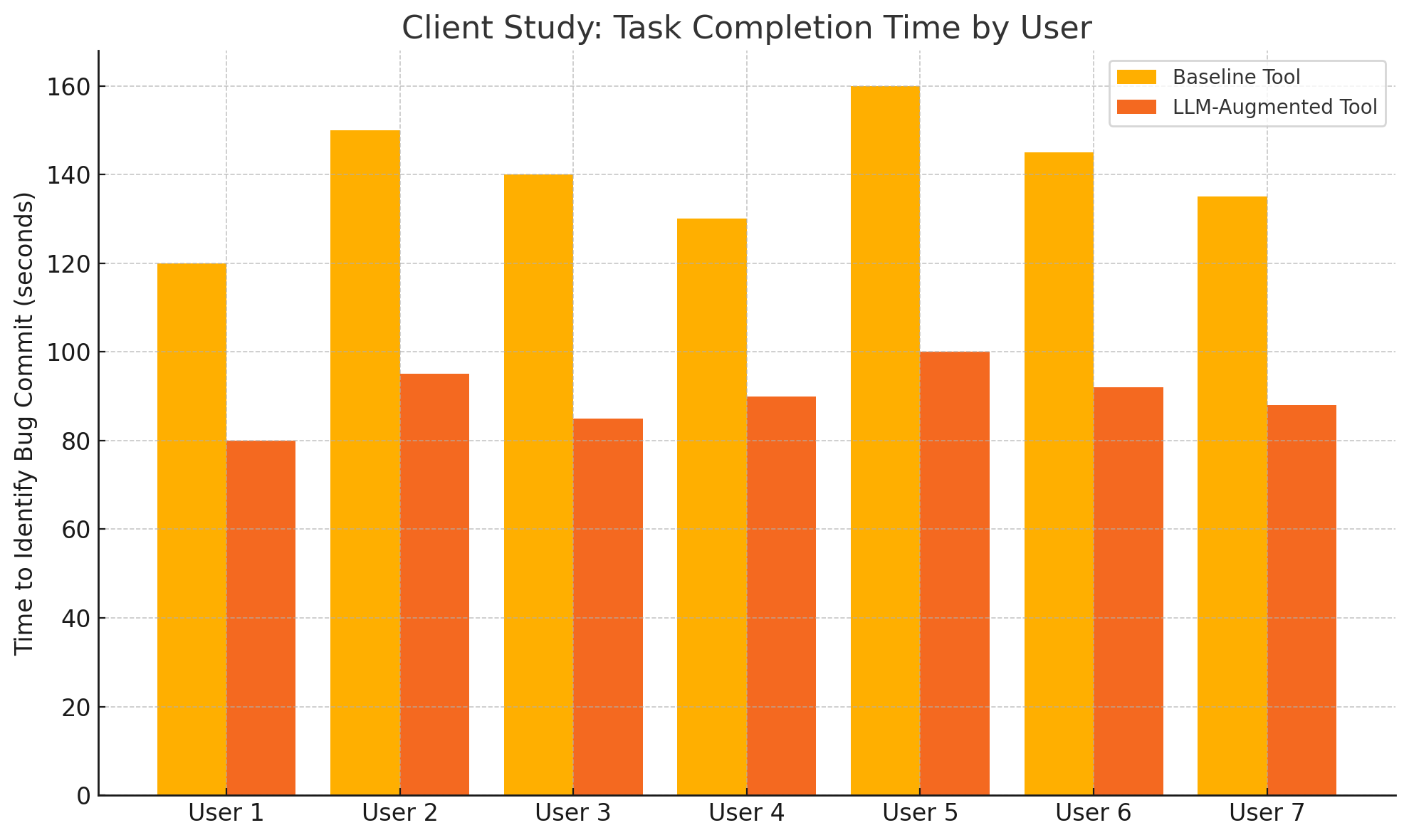}
    \caption{Time saved by the user}
    \label{fig:timesave}
\end{figure}

\section{Conclusion and Future Work}

We introduced a system that augments Git \texttt{bisect} with fine-tuned LLMs and chain-of-thought reasoning for semantic fault localization. Our method improves allows developers to localize semantic changes even under flaky predicates and reduces average debugging time. However, fully automating commit labeling remains semi-decidable and context-dependent. Future work will explore repository-specific pre-finetuning, dynamic prompting strategies, and dataset scaling to improve consistency, inference speed, and model generalizability.

\bibliography{cite}
\bibliographystyle{plain}

\begin{appendices}
    \section{Chain of Thoughts Prompt JSON Schema}
\begin{lstlisting}[language=java,caption={Prompt Schema Json}]
{
  "$schema": "http://json-schema.org/draft-07/schema#",
  "title": "Bisect Sample",
  "type": "object",
  "properties": {
    "target_behaviour": {
      "type": "string",
      "description": "Short description of the behaviour to detect"
    },
    "has_compile_error": {
      "type": "boolean",
      "description": "Whether the code at this commit fails to compile"
    },
    "behaviour_change": {
      "type": "string",
      "description": "Type of behavioural change (e.g., intro, del)"
    },
    "behaviour_confidence": {
      "type": "integer",
      "minimum": 0,
      "maximum": 100,
      "description": "Model confidence score for the detected change"
    },
    "sem_edits": {
      "type": "array",
      "description": "List of semantic edit hypotheses",
      "items": {
        "type": "object",
        "properties": {
          "id": { "type": "string" },
          "kind": { "type": "string" },
          "semantic": { "type": "boolean" },
          "behaviour": { "type": "string" },
          "likelihood": { "type": "integer" },
          "dependency": { "type": "string" },
          "precedent": { "type": "string" }
        },
        "required": ["id","kind","semantic","behaviour","likelihood","dependency","precedent"],
        "additionalProperties": false
      }
    },
    "counterfactual_fix": {
      "type": "string",
      "description": "Suggested fix that would prevent the failing behaviour"
    },
    "reasoning_chain": {
      "type": "array",
      "description": "The step-by-step chain-of-thought of the LLM",
      "items": { "type": "string" }
    },
    "reflection": {
      "type": "string",
      "description": "Self-assessment of model confidence or limitations"
    },
    "bisect_mark": {
      "type": "string",
      "enum": ["good", "bad"],
      "description": "Final binary label assigned by the bisect process"
    }
  },
  "required": [
    "target_behaviour",
    "has_compile_error",
    "behaviour_change",
    "behaviour_confidence",
    "sem_edits",
    "counterfactual_fix",
    "reasoning_chain",
    "reflection",
    "bisect_mark"
  ],
  "additionalProperties": false
}
\end{lstlisting}

\section{Chain of Throughts Prompt Field Explanation}

\subsection*{\texttt{target\_behaviour} (string)}
A short description of the behaviour you’re trying to detect that this code at this snapshot should be able to achieve.

\subsection*{\texttt{has\_compile\_error} (bool)}
Whether the code at this commit fails to compile.

\subsection*{\texttt{behaviour\_change} (intro\,$|$\,del\,$|$\dots)}
The type of behavioral change: introduction, deletion, etc.

\subsection*{\texttt{behaviour\_confidence} (0--100)}
Model’s confidence score for the detected change.

\subsection*{\texttt{sem\_edits} (array of objects)}
A list of semantic edit hypotheses. Each object contains:
\begin{itemize}
  \item \texttt{id} (string): unique edit identifier.
  \item \texttt{kind} (string): the kind of edit (e.g.\ “add”, “modify”).
  \item \texttt{semantic} (bool): whether the edit is semantic.
  \item \texttt{behaviour} (string): behavior label after edit.
  \item \texttt{likelihood} (int): likelihood score of this edit.
  \item \texttt{dependency} (string): dependency context for the edit.
  \item \texttt{precedent} (string): preceding code context.
\end{itemize}

\subsection*{\texttt{counterfactual\_fix} (string)}
A suggested fix that would prevent the failing behavior.

\subsection*{\texttt{reasoning\_chain} (array of strings)}
The LLM’s step-by-step chain-of-thought.

\subsection*{\texttt{reflection} (string)}
A brief self-assessment by the model of its confidence or limitations.

\subsection*{\texttt{bisect\_mark} (good\,$|$\,bad)}
The final binary label assigned to this commit by the bisect process.

\section{User Study}

We recruited seven developers from the functional team based on the following criteria:
\begin{enumerate}
  \item Proficient in programming, with fluency in Python, C\#, and Java.
  \item Employed at the same company and available during standard business hours.
  \item Motivated to learn new technologies.
  \item Unfamiliar with the project’s context and not previously involved with the repository.
  \item Provided with a brief, high-level architectural overview of the codebase.
\end{enumerate}
For our benchmark, we curated a list of bugs and then randomly assigned each participant two distinct tasks—one to debug manually and one using the LLM-enabled bisect. This ensures no participant ever bisects the same bug twice, and that every bug is tested by at least one user under both conditions.
The results of the client study is depicted in figure \ref{fig:timesave}

\begin{figure}
    \centering
    \includegraphics[width=\columnwidth]{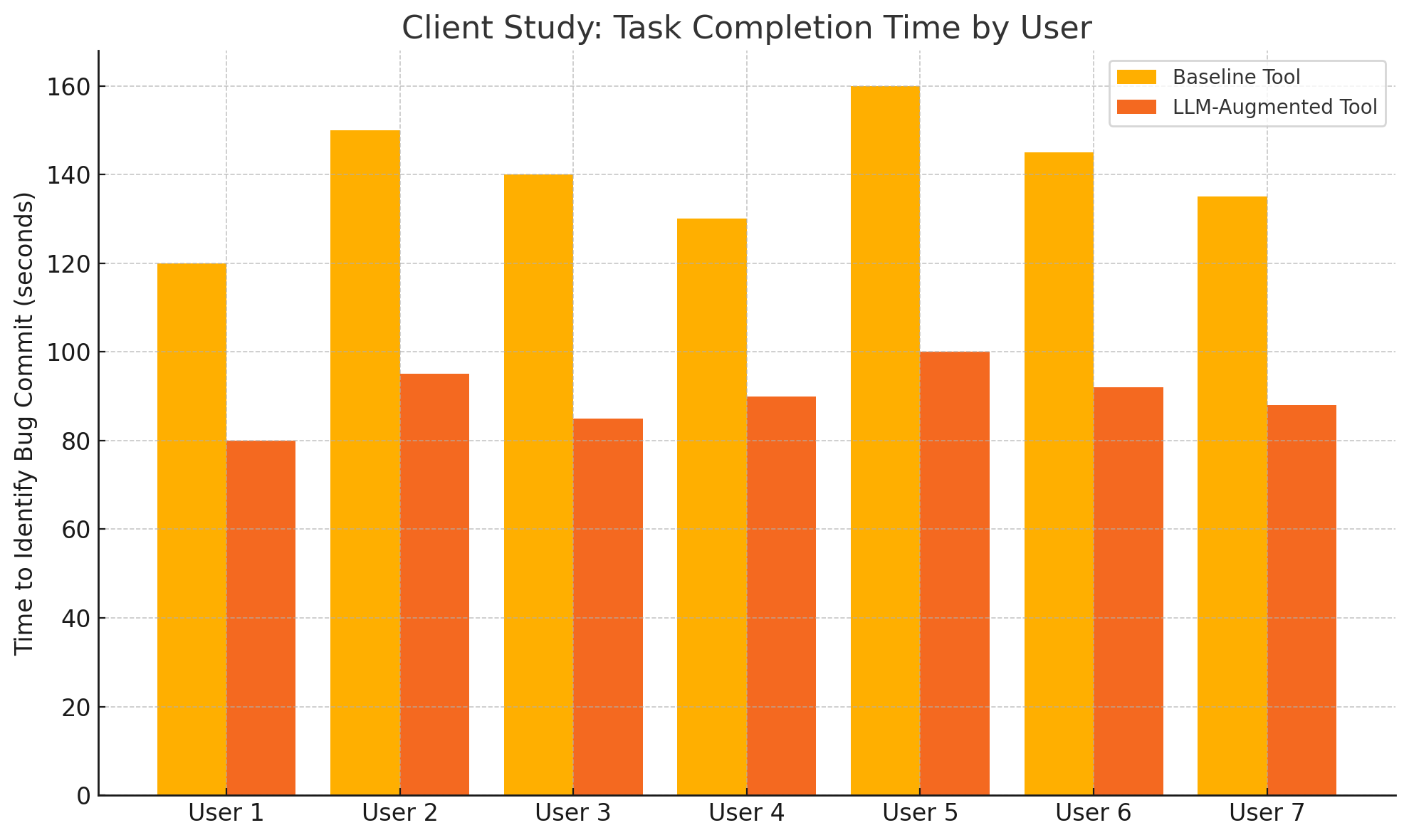}
    \caption{Time saved by the user}
    \label{fig:timesave}
\end{figure}

The total time saved is 45.6\% as shown in the graph above

\section{Bisect Targets by Feature Category}

\begin{itemize}
  \item \textbf{Display / Output Introduction}
    \begin{itemize}
      \item Change welcome banner text
      \item Add ASCII logo above intro
      \item Adjust line spacing in header
      \item Remove trailing newline in greeting
    \end{itemize}

  \item \textbf{Input Handling Introduction}
    \begin{itemize}
      \item Trim whitespace from name field
      \item Reject empty‐string input
      \item Permit multi‐word names
      \item Add “Enter your name:” prompt
    \end{itemize}

  \item \textbf{State‐Transition Logic}
    \begin{itemize}
      \item Split \texttt{INIT → READY} transition
      \item Merge \texttt{READY} and \texttt{PAUSED} states
      \item Change timeout from 5 s to 10 s
      \item Introduce intermediate \texttt{LOADING} state
    \end{itemize}

  \item \textbf{Decision‐Making Rules}
    \begin{itemize}
      \item Update priority rule from A → B
      \item Add fallback when X fails
      \item Swap rule order in \texttt{decide()}
      \item Enforce strict type check in decision
    \end{itemize}

  \item \textbf{Structural Refactor}
    \begin{itemize}
      \item Extract helper into \texttt{utils.py}
      \item Rename \texttt{main.py} → \texttt{app.py}
      \item Move constants to \texttt{config/}
      \item Inline small module into caller
    \end{itemize}

  \item \textbf{Robustness / Error Handling}
    \begin{itemize}
      \item Add \texttt{try/except} around I/O
      \item Raise \texttt{ValueError} on bad input
      \item Log stack trace on failure
      \item Introduce retry loop for network errors
    \end{itemize}

  \item \textbf{Flow-Control / Session Loop}
    \begin{itemize}
      \item Change \texttt{while True} → \texttt{for i in range(5)}
      \item Add break on “quit” command
      \item Refactor loop into run\_session()
      \item Insert sleep(0.1) between iterations
    \end{itemize}

  \item \textbf{Runtime-Launch Safeguard}
    \begin{itemize}
      \item Detect duplicate instances
      \item Lock PID file on start
      \item Abort if config missing
      \item Validate environment variables pre-launch
    \end{itemize}

  \item \textbf{Documentation / Cosmetic}
    \begin{itemize}
      \item Update docstring for \texttt{greet()}
      \item Fix typo in README
      \item Add example usage to docs
      \item Reformat Markdown headers
    \end{itemize}
\end{itemize}

\end{appendices}
\end{document}